\documentclass[preprint,prd,amsmath,amssymb,amsthm,nofootinbib]{revtex4}
\usepackage[mathscr]{euscript}
\usepackage{bm}% bold math
\usepackage{textcomp}
\usepackage{graphicx}% Include figure files
\usepackage{subfigure}
\usepackage{overpic}
\usepackage{multirow}
\usepackage{floatrow}
\usepackage{float} %for fixing the figure
\usepackage{amsmath,amssymb,amsthm}
\usepackage{cases}
\usepackage{cancel}
\usepackage[colorlinks=true,linkcolor=red]{hyperref}
\usepackage{xcolor}
\newfloatcommand{capbtabbox}{table}[][\FBwidth]
\newcommand{\bea}{\begin{eqnarray}}
\newcommand{\eea}{\end{eqnarray}}
\newcommand{\beq}{\begin{equation}}
\newcommand{\eeq}{\end{equation}}

\def\/{\over}

\begin{document}

\title{Quantum Oppenheimer-Snyder primordial black holes as all the dark matter}

\author{
Li-Shuai Wang, and
Xiangdong Zhang\footnote{Corresponding author: scxdzhang@scut.edu.cn}, 
}

\affiliation{
  School of Physics and Optoelectronics, South China University of Technology, Guangzhou 510641, China
}

\begin{abstract}

Primordial black holes (PBHs) are widely considered as candidates for dark matter in many recent studies, and they are often modeled as Schwarzschild or Kerr black holes (BHs), which have curvature singularities. Nevertheless, resolving the classical singularity may require quantum gravity motivated corrections, thereby yielding an effective quantum corrected BH spacetime geometry different from the Schwarzschild or Kerr cases. Therefore, it is well motivated to consider BHs beyond the Schwarzschild or Kerr as viable PBH candidates. Based on these considerations, we investigate quantum Oppenheimer Snyder BHs as PBHs which could account for all the dark matter. Our results show that these BHs have temperatures and greybody factors markedly different from the Schwarzschild case, suppressing Hawking emission and thereby relaxing the $\gamma$-ray constraints from HEAO-1, COMPTEL, and EGRET, which, relative to the Schwarzschild case, broadens the allowed PBH mass window in the asteroid-mass range where PBHs can constitute all of the dark matter.

\end{abstract}

%\pacs{98.80.Cq, 04.50.Kd, 05.70.Fh}

\maketitle
%%%%%%%%%%%%%%%%%%%%%%%%%%%%%%%%%%%%%%%%%%%%%%%%%% 
\section{Introduction}
\label{sec_in}

Black holes (BHs) have attracted tremendous attention in recent years, especially after the first ever BH image was released in 2019~\cite{EventHorizonTelescope:2019dse}, which sparked even greater research interest. This historic achievement not only provided direct evidence for the existence of BHs but also offered a striking test of predictions from general relativity, marking a new stage in our understanding of the Universe. In present theories, BHs can form through two main channels. One class of BHs forms through the gravitational collapse of massive stars at the end of their lives~\cite{Woosley:2002zz,Kalogera:1996ci}. When the compact remnant produced by the collapse exceeds the maximum neutron-star mass, which is of order two to three solar masses, it will undergo further collapse and become a BH. Another possible channel is the formation of BHs in the early Universe from sufficiently large primordial density perturbations. BHs produced via this second channel are known as primordial BHs (PBHs)~\cite{Hawking:1971ei}, and their possible mass range is far broader than that of BHs formed through stellar evolution. PBHs have attracted considerable attention in much recent research, as they may account for a variety of astrophysical and cosmological phenomena including interpretations of some LIGO–Virgo merger signals~\cite{Bird:2016dcv,Clesse:2016vqa,Sasaki:2016jop}, reports of very short-duration microlensing in OGLE observations~\cite{Mroz:2017mvf,Niikura:2019kqi}, and the possibility that PBHs constitute a fraction (or even all) of the dark matter, especially around asteroid-scale masses~\cite{Katz:2018zrn,Barnacka:2012bm,Graham:2015apa,Niikura:2017zjd,Wang:2025aon,Fu:2019ttf}.

However, these PBHs are often modeled as Schwarzschild or Kerr BHs~\cite{villanueva2021brief,arbey2024primordial} which are solutions of general relativity. Although general relativity is highly successful—for example, its predictions for the large scale structure of the Universe are consistent with CMB observations~\cite{aghanim2020planck}, and its prediction of gravitational waves has been confirmed by LIGO~\cite{LIGOScientific:2016aoc}—it cannot explain several fundamental problems, such as spacetime singularities and the BH information paradox~\cite{PhysRevLett.14.57,PhysRevD.14.2460}. These issues suggest that a theory of quantum gravity is ultimately required. In the absence of a complete quantum gravity framework, one can nevertheless study BHs by adopting quantum gravity inspired, quantum corrected BHs and investigating their properties and observational consequences. Compared with Schwarzschild and Kerr BHs, these quantum corrected BHs often give rise to distinct effective spacetime geometries and resolve the singularity problem, thereby modifying the potential barrier around the BH and its temperature, which may in turn lead to different observational constraints~\cite{Calza:2024xdh,Calza:2025mwn}. Therefore, It is worthwhile and well-motivated to explore quantum corrected BHs as PBH candidates. However, at the current stage, the application of loop quantization techniques to BH models lacks a definitive framework—even in the case of the simplest Schwarzschild BH. Nevertheless, significant strides have been made in recent years, particularly within the covariant model and the quantum Oppenheimer-Snyder (qOS) model~\cite{Lewandowski:2022zce,Shi:2024vki,Zhang:2024khj,Chen:2025baz,Belfaqih:2024vfk,Du:2025kcx,Ou:2025bbv,Lin:2024beb,Du:2024ujg}.

Motivated by these considerations, and noting that covariant quantum BHs and regular BHs have already been investigated as possible dark matter candidates in Refs~\cite{Calza:2024fzo,Calza:2025mwn,Calza:2024xdh}, we therefore focus on qOS PBHs~\cite{Lewandowski:2022zce} which are different from Schwarzschild PBHs as candidates for all of the dark matter in this paper. The interior metric of qOS BHs has the same form as the FLRW metric, but the scale factor $a(\eta)$ satisfies a modified Friedmann equation, resulting in a nonsingular interior that undergoes a bounce when the collapsing matter reaches sufficiently small scales. For the exterior metric of the qOS BH, there is an additional term involving the parameter $\alpha$ compared to the Schwarzschild metric. When $\alpha$ is close to zero, the qOS exterior spacetime metric reduces to the Schwarzschild case. Because the spacetime in the qOS case is different from the Schwarzschild case, the greybody factors and the BH temperature may be significantly modified, and these modifications are therefore expected to substantially change the intensity of Hawking radiation and the resulting fraction of dark matter composed of PBHs. Our results show that quantum corrections modify the exterior spacetime of the qOS BH relative to the Schwarzschild case, thereby changing the greybody factors and the BH temperature and greatly affecting the Hawking emission intensity observed at infinity. Moreover, the extragalactic $\gamma$-ray observations from HEAO-1~\cite{Gruber_1999}, COMPTEL~\cite{Schoenfelder:2000bu}, and EGRET~\cite{Strong_2004} provide constraints on Hawking emission that are highly sensitive to such changes, which directly leads to a much broader mass window in which qOS PBHs can constitute all of the dark matter compared with the Schwarzschild case.

The paper is organized as follows: Sec.~\ref{sec:QOS BH} briefly reviews the qOS BH and computes its temperature. Sec.~\ref{sec:greybody factors} discusses how to use the direct method to calculate the greybody factors. Sec.~\ref{sec:Hawking radition} calculates the Hawking radiation of qOS and Schwarzschild BHs. Sec.~\ref{sec:observation constraints} computes and compares the dark matter fractions of Schwarzschild and qOS BHs over a range of masses, considering three different qOS BH scenarios. Finally, Sec.~\ref{sec:conclusions} summarizes the conclusions of this paper. In addition, we set $c=G=\hbar=k_{B}=1$ in this paper.

\section{QOS BH}
\label{sec:QOS BH}

In this section, we briefly review the qOS BH~\cite{Lewandowski:2022zce}, which is obtained by using the Ashtekar–Pawlowski–Singh (APS) scheme of loop quantum cosmology (LQC)~\cite{Ashtekar:2006rx} to modify the classical Oppenheimer–Snyder BH~\cite{Oppenheimer:1939ue}. In this model, the qOS BH interior space-time metric is described by a spatially flat FLRW metric
\begin{equation}\label{eq:dustmetric}
ds^{2}_{APS}=-d\eta^2+a(\eta)^{2}(d\tilde{r}^{2}+\tilde{r}^{2}d\Omega^{2}).
\end{equation}
Where  $d\Omega^2=d\theta^2+\sin^2\theta d\phi^2$, $a(\eta)$ is scale factor, and $(\eta,\tilde{r},\theta,\phi)$ are coordinates. Furthermore, the Friedmann equation which is satisfied by $a(\eta)$ is modified to
\begin{equation}\label{eq:LQCfried}
\begin{aligned} 
H^2=\frac{8\pi}{3}\rho\left(1-\frac{\rho}{\rho_c}\right).
\end{aligned}
\end{equation}
Here, $H$ is the Hubble parameter, $\rho_c=\sqrt{3}/(32\pi^2\gamma^3)$ is the LQC critical density with the Barbero-Immirzi parameter $\gamma$, and the dust energy density is taken to be homogeneous
\begin{equation}
\rho
=\frac{M}{\frac{4}{3}\pi\,\tilde r_0^{\,3}\,a(\tau)^3}.
\end{equation}
where $M$ denotes the total mass of the dust in the APS spacetime and $\tilde r_0$ is the constant comoving radius of the dust boundary, so that the corresponding physical radius is $a(\tau)\tilde r_0$. Eq.~(\ref{eq:LQCfried}) implies that the energy density is bounded as $\rho \le \rho_c$ and reduces to the usual Friedmann equation in the classical regime $\rho \ll \rho_c$. 
Moreover, at $\rho=\rho_c$ one has $H=0$, indicating a turning point that avoids the formation of a singularity in the interior of the qOS BH.

\begin{figure}
    \centering 
    \includegraphics[width=0.7\linewidth]{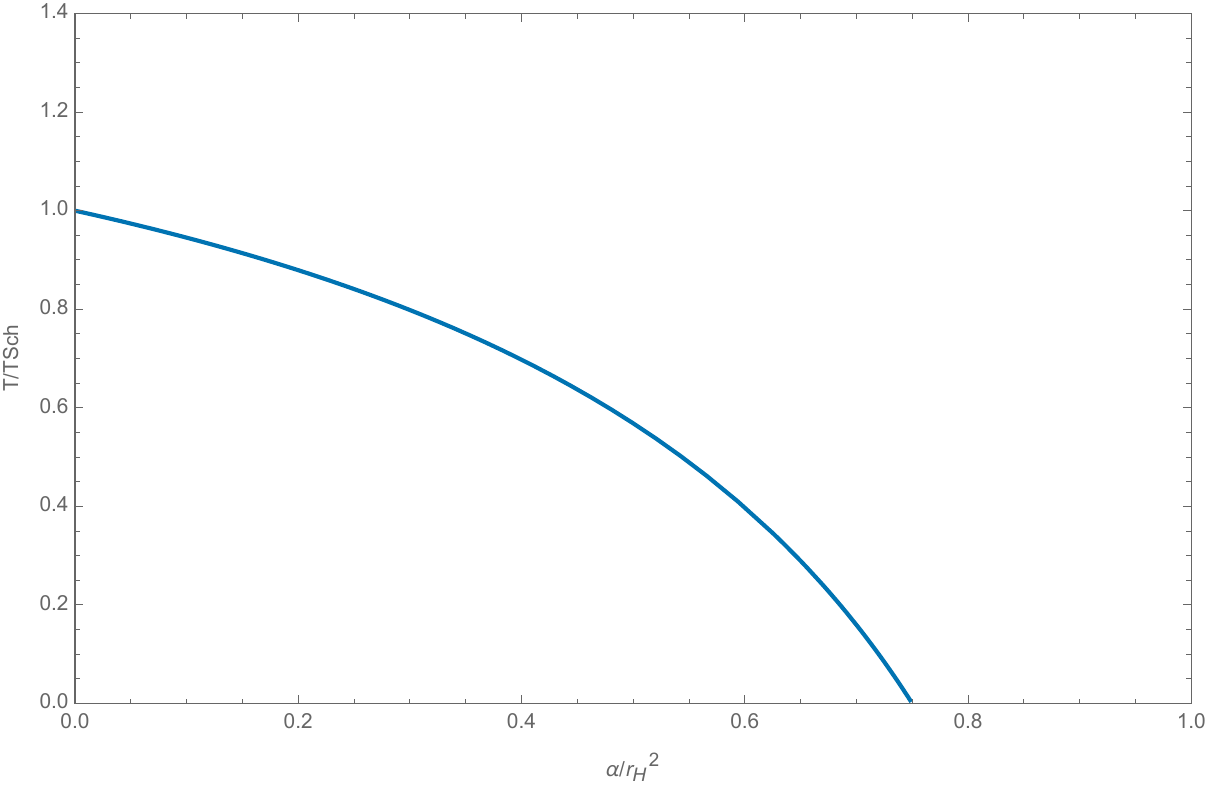}
    \caption{Normalized BH temperature $T/T_{\rm Sch}$ versus $\alpha/r_H^{2}$.
Here $T$ is the temperature of the qOS BH and $T_{\rm Sch}$ is the temperature
of the Schwarzschild BH. For convenience, we set $M=1$ in this figure. }
\label{fig1}
\end{figure}

Now, we discuss the exterior spacetime metric of the qOS BH.  The spacetime outside the dust ball is assumed to be described by a static, spherically symmetric metric
\begin{eqnarray}
ds^2 = -f(r)dt^2 + g(r)^{-1}dr^2 + h(r)d\Omega^2\,.
\label{eq:ds2}
\end{eqnarray}
Using the Israel junction conditions in Refs~\cite{Israel:1966rt,Shi:2024vki}, we can obtain
\begin{eqnarray}
f(r) =g(r)=\left(1-\frac{2M}{r}+\frac{\alpha M^2}{r^4}\right),
\label{eq:fr}
\end{eqnarray}
and
\begin{eqnarray}
h(r) =r^{2}.
\end{eqnarray}
Here, the parameter $\alpha=16\sqrt{3}\pi\gamma^3\ell_p^2$ with the Planck length $\ell_p$. Now we discuss the temperature of the qOS BH, since the temperature is crucial for Hawking radiation. For a static, spherically symmetric BH, the temperature can be calculated by
\begin{eqnarray}\label{temperature}
T=\sqrt{\frac{g(r)}{f(r)}}\frac{f'(r)}{4\pi}\vert_{r_H}\,,
\label{eq:temperature}
\end{eqnarray}
where $r_{H}$ is the horizon radius. Using Eq.~(\ref{temperature}) in natural units, the temperature of the Schwarzschild BH $T_{Sch}=1/8\pi M$ can be obtained. We plot the ratio $T/T_{\rm Sch}$ versus $\alpha/r_{H}^{2}$ in Fig.~\ref{fig1} and we find that the temperature of the qOS BH is lower than that of the Schwarzschild case for $\alpha /r_{H}^{2}>0$. Moreover, when $\alpha/ r_{H}^{2}$ is about equal to 0.75, the temperature of the qOS BH will be close to zero.

\section{greybody factors}
\label{sec:greybody factors}

It is known that the Hawking radiation is not blackbody radiation at infinity, due to BHs inducing a potential barrier around them. Particles emitted by the BH have a certain probability of tunneling through this potential barrier and this transmission probability is known as the greybody factor. In other words, different BHs may induce different potential barriers, and the potential barriers may significantly affect the probability of tunneling of the particles, so it is necessary to check the greybody factors of the qOS BH and the Schwarzschild BH.  In this section, we will use the method in Refs~\cite{Calza:2024fzo,Calza:2024xdh} to calculate the greybody factors. The greybody factors are obtained by solving the Teukolsky equation, and the Teukolsky equation for the evolution of massless perturbations $\Upsilon_s$ of spin $s$ in the static spherically symmetric spacetime is described by~\cite{Arbey_2021}
\begin{align} 
& - \frac{h}{f}\partial^2_t \Upsilon_s + s \sqrt{\frac{g}{f}}\left( \frac{hf'}{f} - h' \right ) \partial_t \Upsilon_s \nonumber \\ \nonumber
&+ g h \partial_r^2 \Upsilon_s + \left ( \frac{hg'}{2} + (s+1/2) \frac{ghf'}{f}+(s+1)gh' \right ) \partial_r \Upsilon_s \\ \nonumber
&+ \left ( \frac{1}{\sin{\theta}} \partial_\theta (\sin{\theta}\;\partial_\theta )+\frac{1}{\sin^2{\theta}} \partial_\phi^2 \right.\\ \nonumber
&  \left. + \frac{2is \cot{\theta}}{\sin{\theta}} \partial_\phi - s^2 \cot^2{\theta}-s\right ) \Upsilon_s \\ \nonumber
& +\left ( s \frac{hgf''}{f} +\frac{3s-2s^2}{4}(2gh''+g'h') + \frac{s}{2} (\frac{hg'f'}{f} -\frac{ghf'^2}{f^2}) \right. \\ 
& \left. +\frac{2s^2-s}{4} \frac{gh'^2}{h} + \frac{2 s^2 +5s}{4} \frac{g f' h'}{f} \right ) \Upsilon_s = 0\,,
\label{eq:teukolsky}
\end{align}
where $(t,r,\theta,\phi)$ are standard spherical coordinates, a prime denotes a derivative with respect to $r$, and the massless perturbations $\Upsilon_s$ of spin $s$ can be separated as follows
\begin{eqnarray}
\Upsilon_s= \sum_{l,m} e^{-i \omega t } S^{s}_{l,m}(\theta,\phi) R_s(r)\,.
\label{eq:upsilon}
\end{eqnarray}
Here, $l$ is the angular node number, $m$ is the azimuthal node number, and $\omega$ is the perturbation frequency. In addition, the function $S^s_{l,m}$ satisfies the equation as follows~\cite{Fackerell:1977shn} 
\begin{align}
&\left ( \frac{1}{\sin\theta}\partial_\theta(\sin\theta\,\partial_\theta)+\csc^2\theta\,\partial_\phi^2 \right. 
\left. + \frac{2is\cot\theta}{\sin\theta}\partial_\phi+s-s^2\cot^2\theta+\lambda_l^s \right ) S_{l,m}^s=0\,,
\label{eq:slms}
\end{align}
where $\lambda^s_l = l(l+1)-s(s+1)$ is the separation constant which is determined by $l$ and $s$. Based on above discussion, the Teukolsky equation~(\ref{eq:teukolsky}) for the radial component of a field with spin $s$, $R_{s}$, in the static spherically symmetric spacetime is described by~\cite{Arbey_2021}
\begin{eqnarray}\label{Teukolsky1} 
A_s(B_s R'_s)'+ \left [ \frac{h}{f}\omega^2+i\omega s\sqrt{\frac{g}{f}} \left ( h'-\frac{h f'}{f} \right ) + C_s \right ] R_s=0\,. \nonumber \\
\label{eq:radialteukolsky}
\end{eqnarray}
Where
\begin{eqnarray}
A_s= \sqrt{\frac{g}{f}} \frac{1}{(f h)^s}\,,
\label{eq:as}
\end{eqnarray}
\begin{eqnarray}
B_s=\sqrt{f g } (f h)^sh\,,
\end{eqnarray}
\begin{align}
C_s &= s \frac{g h f''}{f} + \frac{s}{2} \left ( \frac{h g' f' }{f} - \frac{g h f'^2}{f^2} \right ) \nonumber \\ 
&+ \frac{s(3-2s)}{4} \left( 2 g h'' + g' h'  \right) +\frac{s(2s-1)}{4} \frac{g h'^2}{h}\nonumber \\ 
&+\frac{s(2s+5)}{4}\frac{g f' h'}{f}-\lambda^s_l-2s\,,
\end{align}
To calculate greybody factors, we use the shooting method, and the rescaled coordinate $x$ is defined by
\begin{eqnarray}
x\equiv\frac{r-r_H}{r_H}\,,
\label{eq:varx}
\end{eqnarray}
where we set the horizon radius $r_{H}=1$ for convenience. Based on above discussions, Eq.~(\ref{Teukolsky1}) can be rewritten as follows
\begin{eqnarray}\label{Teukolsky2}
\mathcal{A}_s\ddot{R}_s+\mathcal{B}_s\dot{R}_s+\mathcal{C}_sR_s=0\,.
\end{eqnarray}
Here 
\begin{align}
\mathcal{A}_s=gA_{s}B_{s}, 
\end{align}
\begin{align}
\mathcal{B}_s=gA_{s}\dot{B}_s, 
\end{align}
\begin{align}
\mathcal{C}_s= g\left [ \frac{h}{f}\omega^2+i\omega s\sqrt{\frac{g}{f}} \left ( \dot{h}-\frac{h \dot{f}}{f} \right ) + C_s \right ],
\end{align}
and an overdot denotes a derivative with respect to $x$. For consistency with the Teukolsky equation which is rewritten as Eq.~(\ref{Teukolsky2}) and calculating greybody factors, the spacetime metrics need to be rewritten. For the qOS BH case, we can obtain
\begin{align}\label{fx}
f(x)=g(x)=1-\frac{1}{(1+x)^4}+\frac{2x(3+3x+x^2)(-1+\sqrt{1-\alpha})}{(1+x)^4\alpha},
\end{align}
and
\begin{align}\label{hx}
h(x)=(1+x)^2.
\end{align}

For calculating greybody factors, it is necessary to obtain the radial function $R_{s}$ at infinity. However, the $R_{s}$ cannot be obtained at infinity immediately. Instead, we can obtain the numerical solution of the Teukolsky equation near the horizon, and then the $R_s$ at infinity can be obtained by solving Eq.~(\ref{Teukolsky2}) numerically. For this purpose, $R_s$ needs to be obtained around the horizon and it takes the form as follows~\cite{Rosa:2012uz,Rosa:2016bli}
\begin{eqnarray}\label{Rxs}
R_s(x)=x^{q}\sum_{n=0}^\infty a_n x^n\,.
\end{eqnarray}
By combining Eqs.~(\ref{fx})--(\ref{Rxs}) and setting $a_{0}=1$, we expand Eq.~(\ref{Teukolsky2}) to the lowest order near the horizon, and obtain 
 \begin{eqnarray}
          q(q-1) \tau^2  + q (s+1) \tau^2 + \omega (\omega - i s \tau)=0\,,
          \label{Eu-Cau}
        \end{eqnarray}
whose solutions are
    \begin{eqnarray}\label{q}
          q_1=-s - \frac{i \omega}{\tau} \;\; ,\;\;   q_2=\frac{i \omega}{\tau},  
    \end{eqnarray}
where 
\begin{eqnarray}
\tau=4+\frac{6(-1+\sqrt{1-\alpha})}{\alpha}.
\end{eqnarray}
Moreover, we equivalently assume that particles are incident from spatial infinity, cross the potential barrier, and fall into the BH, so we select the ingoing mode at the horizon, which corresponds to choosing the first solution for $q$. Therefore, to further determine $R_s$ near the horizon, we substitute Eq.~(\ref{Rxs}) (with $q=q_1$) into Eq.~(\ref{Teukolsky2})
and expand Eq.~(\ref{Teukolsky2}) order by order (from the lowest order upward) to determine the coefficients $a_n$ recursively. In this way, all the coefficients in Eq.~(\ref{Rxs}) are obtained. The near-horizon solution in Eq.~(\ref{Rxs}) is employed to impose the boundary conditions and to numerically solve the radial equation out to large distances, where the solution takes the general form~\cite{Rosa:2016bli}
\begin{align}
R_s(x)\xrightarrow{r\to\infty}\;
{}_sR^{lm}_{\mathrm{in}}\frac{e^{-i\omega x}}{x}
+{}_sR^{lm}_{\mathrm{out}}\frac{e^{i\omega x}}{x^{2s+1}}\,.
\label{eq:far}
\end{align}
The greybody factors can then be computed from the coefficient ${}_sR^{lm}_{\rm in}(\omega)$ and can be described by
\begin{eqnarray}\label{greybody factors}
\Gamma^s_{l m}(\omega)=\delta_s \vert _s R^{l m}_{\text{in}}(\omega)\vert^{-2}\,,
\label{eq:gamma}
\end{eqnarray}
where 
\begin{eqnarray}\label{deltas}
\delta_s = p\tau \frac{ i e^{i \pi s} (2 \omega)^{2s-1} \Gamma \left ( 1-s- \frac{2 i \omega}{\tau} \right ) }{\Gamma \left ( s-\frac{2 i \omega}{\tau} \right ) } \,.
\label{eq:delta}
\end{eqnarray}

\begin{figure}
    \centering
    \includegraphics[width=0.7\linewidth]{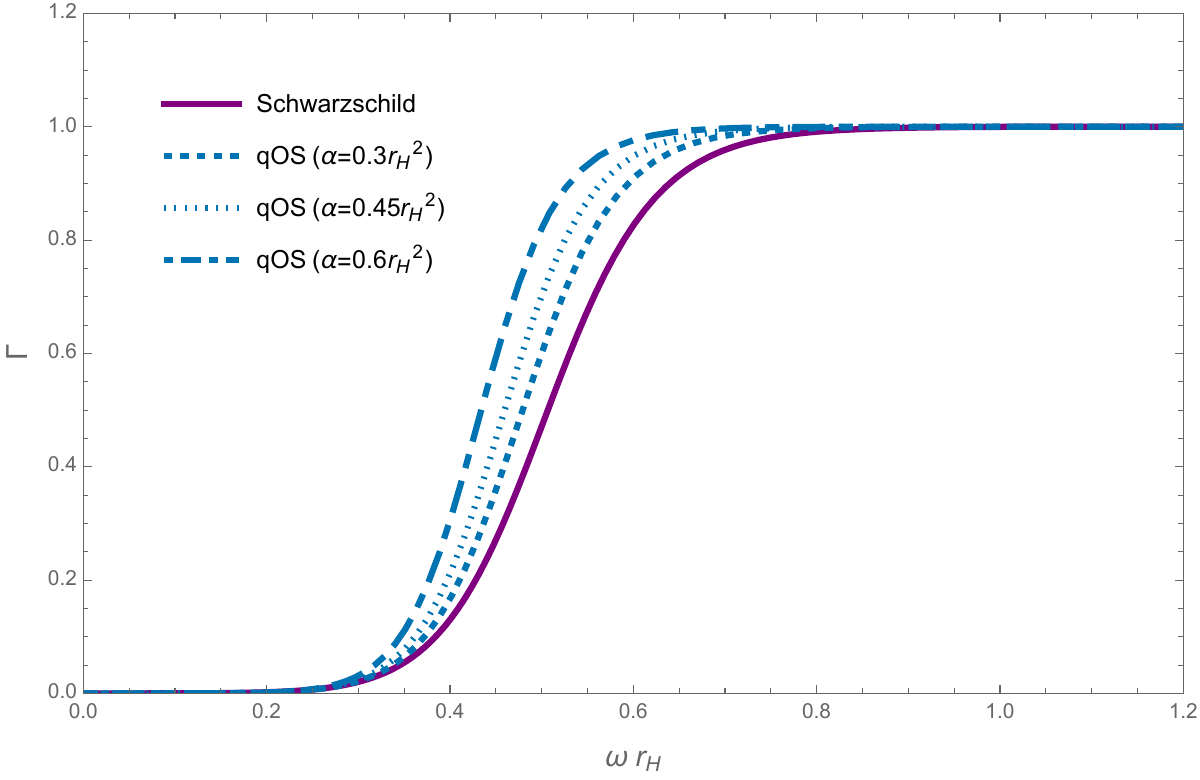}
    \caption{Greybody factors are shown as a function of $\omega r_H$. In this plot, we set $s=1$ and $l=1$. Moreover, the solid purple line corresponds to the Schwarzschild case, while the dashed, dotted, and dot-dashed blue lines correspond to qOS BHs with $\alpha=0.3r_{H}^2$, $\alpha=0.45r_{H}^2$, and $\alpha=0.6r_{H}^2$, respectively.}
    \label{greybody}
\end{figure}

\noindent Here, $p$ and $\tau$ depend on the metric. In the Schwarzschild and qOS BH cases, $p=1$, since these metrics are $t$--$r$ symmetric. It is worth noting that the modes with different $m$ for a given $l$ are degenerate in a spherically symmetric spacetime. In this way,
we obtain ${}_sR^{lm}_{\text{in}}(\omega)$ numerically, and then combine Eqs.~(\ref{greybody factors}) and~(\ref{deltas}) to calculate the greybody factors. In addition, we set $s=1$ and $l=1$ in Fig.~\ref{greybody} which shows the Schwarzschild case and the qOS BH cases for $\alpha=0.3r^{2}_{H}$, $\alpha=0.45r^{2}_{H}$, and $\alpha=0.6r^{2}_{H}$. It is obvious that the greybody factors in the qOS BH case are larger than those in the Schwarzschild case in certain frequency ranges. This means that, in the qOS BH case, photons can more easily penetrate the gravity-induced potential barrier and reach infinity in the frequency range where the greybody factor exceeds that of the Schwarzschild case.

\section{Hawking radiation}
\label{sec:Hawking radition}
Based on the above discussion, we have already obtained the temperature and the greybody factors in the qOS BH case. The lower temperature reduces the Hawking radiation and larger greybody factors enhance the Hawking radiation. Therefore, it is necessary to evaluate the total impact of both the temperature and the greybody factors on the Hawking radiation. In this section, we focus on the primary photon spectrum, and the Hawking radiation is given by~\cite{Hawking:1975vcx,Page:1976df}
\begin{eqnarray}\label{prim}
\frac{d^2N_i}{dtdE_i}=\frac{1}{2\pi}\sum_{l,m}\frac{n_i\Gamma^s_{l,m}(\omega)}{ e^{\omega/T}\pm 1}\,,
\label{eq:d2ndtdei}
\end{eqnarray}
where the index $i$ labels the particle species, $n_i$ is the number of internal degrees of freedom of that species, and $N_i$ denotes the number of emitted particles of species $i$. For photons, $n_i=n_{\gamma}=2$, corresponding to the two physical polarization states. In addition, $t$ is the asymptotic coordinate time, and in natural units $\omega=E_i$. The azimuthal node number $m$ in greybody factors is degenerate for spherical symmetry spacetimes, so Eq.~(\ref{prim}) can be written as
\begin{eqnarray}\label{prim2}
\frac{d^2N_i}{dtdE_i}=\frac{1}{2\pi}\sum_{l}(2l+1)\frac{n_i\Gamma^s_{l}(\omega)}{ e^{\omega/T}\pm 1}\,.
\label{eq:d2ndtdei}
\end{eqnarray}

\begin{figure}[h]
\centering
\includegraphics[width=0.65\linewidth]{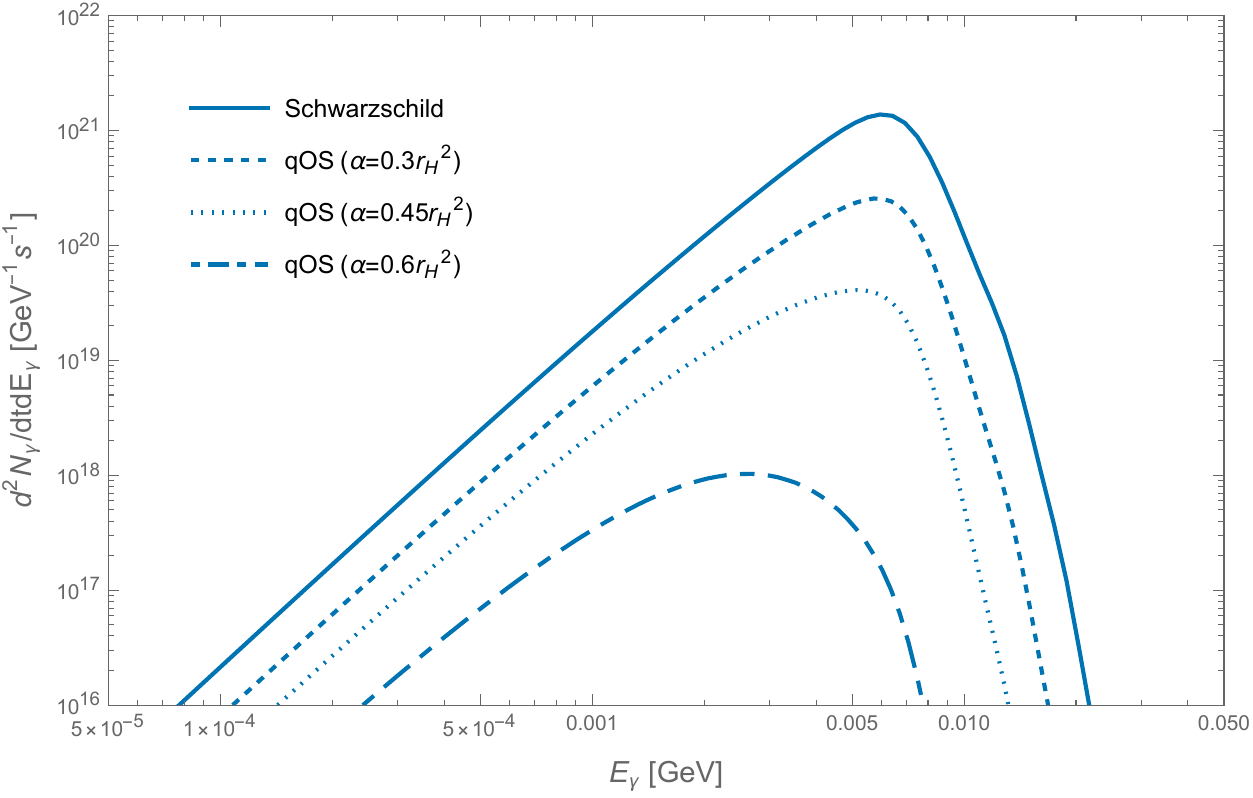}
\caption{\label{Hawking radiation} The particle number density of the Hawking radiation as a function of particle energy. We fix the mass of PBH $M_{pbh}=10^{16}\,\mathrm{g}$ and show the results. The solid line is the Schwarzschild case, the dashed line is the qOS case for $\alpha=0.3r_{H}^{2}$, the dotted line is the qOS case for $\alpha=0.45r_{H}^{2}$ and the dotted dashed line is the qOS case for $\alpha=0.6r_{H}^{2}$.  }
\end{figure}

Using Eq.~(\ref{prim2}), we compute the Hawking radiation for the Schwarzschild case and the qOS case with different values of $\alpha$, and present the results in Fig.~\ref{Hawking radiation}. In the numerical evaluation, we sum over the angular-momentum modes up to $l=4$. As can be seen, for the qOS PBH the particle number density of Hawking radiation is smaller than that of the Schwarzschild BH, and it decreases further as $\alpha/r_{H}^{2}$ increases. Although the expression indicates that the particle number density is affected by both the BH temperature and the greybody factors, and the greybody factors in the qOS case are larger than the greybody factors in the Schwarzschild case in the relevant frequency range, the overall emission is still suppressed, showing that the temperature effect plays the dominant role.

\section{observation constraints}
\label{sec:observation constraints}

Based on the discussion above, we have obtained the Hawking radiation spectrum. We now use the extragalactic $\gamma$-ray background together with our theoretical Hawking-radiation prediction to constrain the PBH number density ($n_{pbh}$), and hence determine the fraction of dark matter that can be accounted for by PBHs of different masses. It is worth noting that Schwarzchild PBHs with masses below $5\times10^{14}$ g may have already evaporated, while PBHs with masses over $10^{15}\,\mathrm{g}$ can lose a negligible fraction of their mass through evaporation (compared to the total mass of an individual PBH)~\cite{Calza:2024fzo,Page:1976df}. Therefore, we restrict our analysis to PBHs with $M_{\rm pbh}\gtrsim 10^{15}\,\mathrm{g}$ for simplicity. The extragalactic $\gamma$-ray background photon flux is given by~\cite{Gruber_1999,Schoenfelder:2000bu,Strong_2004,Carr:2009jm} 
\begin{eqnarray}
I(E_{\gamma 0}) \equiv \frac{c}{4\pi}n_{\gamma 0}(E_{\gamma 0})\,,
\label{flux}
\end{eqnarray}
where $E_{\gamma0}$ is the energy of photon, and the present number density of photons $n_{\gamma0}(E_{\gamma0})$ is described by
\begin{align}
\begin{split}
&n_{\gamma 0}(E_{\gamma 0})\\
&= n_{\text{pbh}}(t_0) E_{\gamma 0}\int^{t_0}_{t_{\star}} dt(1 + z) \frac{d^2{N}_{\gamma}}{dt dE_{\gamma}}(M_{\text{pbh}},(1+z)E_{\gamma 0})\\
& = n_{\text{pbh}}(t_0) E_{\gamma 0}\int^{z_{\star}}_{0} \frac{dz}{H(z)}\frac{d^2{N}_{\gamma}}{dt dE_{\gamma}}(M_{\text{pbh}},(1+z)E_{\gamma 0})\,.
\label{npbh1}
\end{split}
\end{align}
Here, $t_0$ is the present time, $t_{\star}$ is the cosmic time, $z_{\star}$ is the redshift at recombination, and $H(z)$ is the Hubble parameter. Combining Eq.~(\ref{flux}) and Eq.~(\ref{npbh1}), we can obtain $n_{\rm pbh}(t_0)$ for each $M_{\rm pbh}$. Then, the fraction of PBHs in dark matter can be calculated by
\begin{eqnarray}
f_{\text{pbh}}(M_{\text{pbh}}) \equiv \frac{\Omega_{\text{pbh}}}{\Omega_{\text{dm}}} = \frac{n_{\text{pbh}}(t_0)M_{\text{pbh}}}{\rho_{\text{crit},0}\Omega_{\text{dm}}}\,,
\label{eq:npbhtofpbh}
\end{eqnarray}

\begin{figure}[h]
\centering
\includegraphics[width=0.7\linewidth]{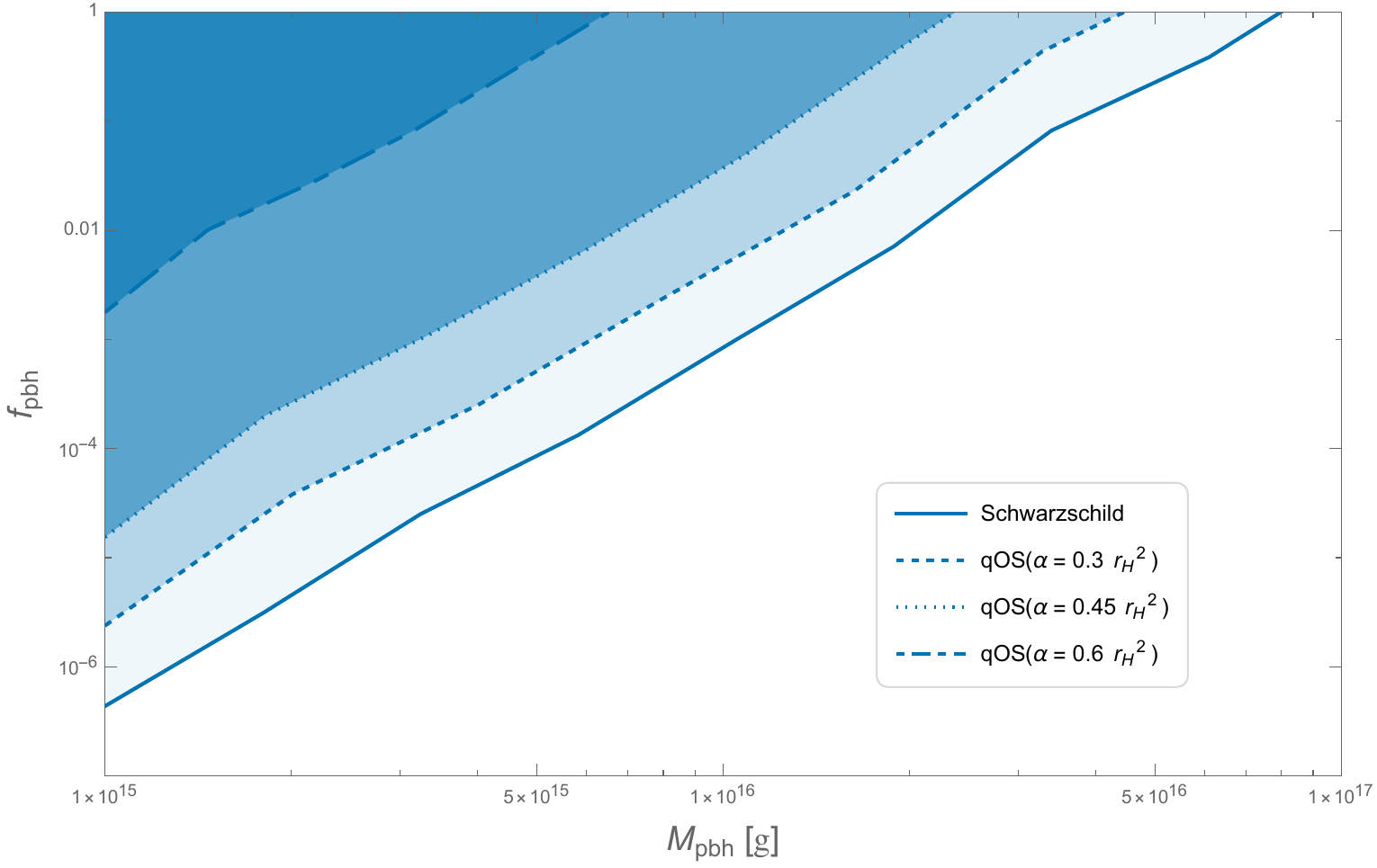}
\caption{The dark-matter fraction of PBHs as a function of PBH mass. The solid line corresponds to the Schwarzschild case. The dashed, dotted, and dot-dashed lines correspond to qOS PBHs with different values of $\alpha$, respectively.\label{fpbh} }
\end{figure}

\noindent where $\rho_{\text{crit},0}=3H_0^2/8\pi G$ is the present-day critical density, and $H_0$ is the present-day Hubble parameter. In Fig.~\ref{fpbh}, we plot the dark-matter fraction for Schwarzschild and qOS PBHs for different values of $\alpha$. We find that the observational constraints on the dark-matter fraction of qOS PBHs are weaker than those for Schwarzschild PBHs. Especially, the mass range in which qOS PBHs for $\alpha=0.6r^2_H$ case can account for all of the dark matter is enlarged by more than an order of magnitude compared with the Schwarzschild case.

 \section{conclusions}
 \label{sec:conclusions}
 PBHs are often considered as Schwarzschild or Kerr BHs in many studies, which are solutions of general relativity, but there are problems (such as the singularity problem) that may not be fully addressed within general relativity, indicating the potential need for quantum gravity effects. Although we have not yet developed a complete theory of quantum gravity that can resolve spacetime singularities and fully compute Hawking radiation, we can still investigate these issues using quantum corrected BHs and the effective spacetime geometries they induce. Due to quantum corrections, the temperatures and greybody factors of quantum-corrected BHs may differ from those of Schwarzschild and Kerr BHs. This can in turn modify the intensity of Hawking radiation and thereby affect observational constraints on the fraction of dark matter in asteroid-mass PBHs. In this paper, we investigate qOS BHs which are non-singular as candidates for constituting all of the dark matter, and compare them with the Schwarzschild case. We found that the greybody factors were larger than those in the Schwarzschild case in key frequency ranges. This indicates that photons can more easily penetrate the gravity-induced potential barrier around the BH, thereby enhancing the Hawking radiation. Meanwhile, the temperature of qOS BH is lower than that in the Schwarzschild case, and the low temperature indicates that the Hawking radiation will be weaker. We incorporated the influences of greybody factors and temperature and plotted the Hawking radiation in Fig.~\ref{Hawking radiation}, while fixing the PBH mass $M=10^{16}\mathrm{g}$ and summing $l$ from one to four. It is obvious that Hawking radiation is weaker than that in the Schwarzschild case. Moreover, one can clearly see that if $\alpha /r_{H}^{2}$ is larger, the Hawking radiation of the qOS BH will be weaker. In particular, for the qOS case with $\alpha/r_{H}^{2}=0.6$, the Hawking radiation is more than three orders of magnitude smaller than that in the Schwarzschild case. These results showed that the temperature dominates the Hawking radiation. Then, we combined the observational data of 1$\sigma$ upper limits on the extragalactic $\gamma$-ray background flux and the Hawking radiation which we obtained by calculating to constrain the fraction of dark matter in PBHs. We finally obtained the constraints on the fraction of dark matter in PBHs for different PBH masses. In Fig.~\ref{fpbh}, it is obvious that the allowed mass windows in the qOS BH cases are broader than those in the Schwarzschild case. In particular, for $\alpha/r_{H}^{2}=0.6$, the allowed windows are broadened by more than one order of magnitude compared with the Schwarzschild case.

\begin{acknowledgments}

This work is supported by National Natural Science Foundation of China (NSFC) with Grants No.12275087.

\end{acknowledgments}

\bibliographystyle{JHEP}
\bibliography{ref}

\end{document}